\documentclass{article}

\usepackage{enumitem}

\usepackage[preprint]{neurips_2023}

\usepackage[utf8]{inputenc} 
\usepackage[T1]{fontenc}  
\usepackage[hidelinks]{hyperref} 
\usepackage{url}  
\usepackage{booktabs} 
\usepackage{amsfonts} 
\usepackage{nicefrac} 
\usepackage{microtype}  
\usepackage{xcolor} 

\usepackage{rotating}

\usepackage{graphicx}
\usepackage{multicol}
\usepackage{wrapfig}
\usepackage{arydshln}
\usepackage{caption, subcaption}
\usepackage[normalem]{ulem}

\newcommand{\shd}[1]{{\color{magenta}}}
\title{Test-Time Training for Speech}

\author{%
  Sri Harsha Dumpala \\
  Vector Institute and Dalhousie University \\
  \And
  Chandramouli Sastry \\
  Vector Institute and Dalhousie University \\
  \And
  Sageev Oore \\
  Vector Institute and Dalhousie University \\
}

\begin{document}

\maketitle

\begin{abstract}
In this paper, we study the application of Test-Time Training (TTT) as a solution to handling distribution shifts in speech applications. In particular, we introduce distribution-shifts to the test datasets of standard speech-classification tasks---for example, speaker-identification and emotion-detection---and explore how Test-Time Training (TTT) can help adjust to the distribution-shift. In our experiments that include distribution shifts due to background noise and natural variations in speech such as gender and age, we identify some key-challenges with TTT including sensitivity to optimization hyperparameters (e.g., number of optimization steps and subset of parameters chosen for TTT) and scalability (e.g., as each example gets its own set of parameters, TTT is not scalable). Finally, we propose using BitFit -- a parameter-efficient fine-tuning algorithm proposed for text applications that only considers the bias parameters for fine-tuning -- as a solution to the aforementioned challenges and demonstrate that it is consistently more stable than fine-tuning all the parameters of the model. 

\end{abstract}

\section{Introduction}
\label{intro}
 
Deep learning methods achieve impressive results in a variety of speech-based downstream tasks when the train and test data are in-distribution \citep{Gulati_conformer, snyder2017deep, zou2022speech}.
In practice, however, the train and test distributions are usually different, i.e., there exists a distributional shift between the train and test data. 
In speech, such distributional shifts can be introduced due to inter-speaker variations such as speaking style, gender, age, etc., or due to background induced noises such as babble, living room, traffic, etc. 
These distributional shifts significantly degrade the performance of the deep learning models \citep{ASR_benchmarking, garcia2019speaker, cross_corpus_emotion}.
In real-world applications, some form of distributional shifts often occurs in the test data, making it of vital importance for deep learning models to be robust to these shifts.


One approach to handling distributional shift is with \textit{train-time} techniques~\citep{domainAdapt_ASR, adversarial_spk_rec}, in which we need to anticipate the type of distributional shifts that can occur during testing, and then train the model on data collected with this anticipated list of distributional shifts. In practice, the anticipated list of distributional shifts is non-exhaustive, and there is no guarantee that the trained model can generalize well to an unseen domain at test-time.

Another
interesting approach which attained significant improvements in performance for imaging tasks is test-time training (TTT) \citep{sun2020test, liu2021ttt++, ttt_mae}. In TTT, we update the model at inference using the test-sample. As the test sample does not have a label, a self-supervised learning task is used for this update. 

The efficacy of TTT is impacted by the choice of the self-supervised learning task \citep{liu2021ttt++}. \citet{ttt_mae} shows that masked auto-encoding on images is a suitable task for TTT.
Motivated by the success of the transformer-based masked autoencoders (MAE) for speech \citep{mae_audio_neurips}, we extend a test-time training approach based on MAE \citep{ttt_mae} to speech in this work. To the best of our knowledge, this is the first work to adapt TTT to the speech domain. We show that TTT-MAE for speech shows significant improvements on three different downstream tasks under a variety of distributional shifts. 

Two major challenges of using TTT during inference are maintaining stable performance robust to (reasonable) ranges of hyperparameters, and the potential high computational cost, which is due to both (a) increased memory requirements for updating all parameters of the model, and (b) inability to process a batch of samples if the individual samples in the batch are associated with different distributional shifts. We show that, for speech, it is possible to make significant improvements with respect to all of these issues, i.e. we improve stability, reduce memory requirements, and allow batch processing, all by using parameter-efficient training.
Specifically, we show that using bias fine-tuning \citep{2022bitfit}, we can process a batch of test samples, even under the condition that each test sample has a different distributional shift.

\section{Related work}
\label{rel_work}

\textbf{Domain adaptation and generalization.}
These methods are based on the assumption that  models will have access to labelled data from the train distribution and unlabelled (labelled) data from the test distribution. A common strategy in domain adaptation is to learn domain invariant features between train and test data distributions \citep{sun2016deep, tzeng2017adversarial, long2018conditional}. Another approach is to perform self-training on the test distribution by generating pseudo-labels for the unlabelled data \citep{xie2020self}. 
Domain generalization techniques mainly resort to adversarial training, meta learning or adversarial data augmentation \citep{yang2021adversarial, balaji2018metareg, dou2019domain, volpi2018generalizing}.
In speech, domain adaptation and generalization techniques have been applied to tasks such as automatic speech recognition, emotion recognition and speaker classification \citep{domainAdapt_ASR, hu2021redat, MMD_emo, adversarial_spk_rec}. All these methods assume information about the test domain is available during training. \\

\textbf{Learning at inference.}
In the above techniques, the model is trained to generalize to all possible distributional shifts. But anticipating every possible distributional shift at train time is not feasible, particularly in real world applications. Most models trained using domain generalization techniques are fixed during inference even when the test distribution changes. To alleviate this problem, another line of work is to adapt the model to  test samples at inference. Methods to update the model at inference can be classified into two types: test-time adaptation and test-time training. 

\textit{Test-time Adaptation (TTA):} These methods allow using off-the-shelf models without any additional training. In general terms, test-time adaptation focuses on adapting models that were not trained with a special configuration prior to being used at inference \citep{TTA_NIPS_2022, TTT_audio_class_2023}. One of the first approaches of this category, called TENT \citep{wang2021tent}, requires the model and target data. It then updates the model layers by minimizing the Shannon entropy of predictions. \citet{mummadi2021test} improves TENT by using a log-likelihood ratio instead of entropy, and by estimating target batch statistics. Another approach is to update batch normalization (BN) statistics using large number of test samples \citep{nado2020evaluating, schneider2020improving}. 
SITA \citep{khurana2021sita} is one such approach which can be used on a single test data example. SITA generates a pseudo-batch by randomly augmenting this example and then computes  statistics on this pseudo-batch.
TTA in computer vision is heavily targeted on the BN layer’s adaptation by re-estimating batch statistics on target data. In this work we use transformer models, which have achieved state-of-the-art performance on many speech-based downstream tasks. Since transformer models are not equipped with BN layers, as the length of batched input sequences are different, TTA techniques cannot be applied directly to speech. Only one work has applied TTA to speech \citep{TTT_ASR_2023}; this approach appears limited  to ASR models trained with CTC loss.

\textit{Test-time training (TTT)}: The basic paradigm in TTT ~\citep{sun2020test} is to use a test-time task (usually a self-supervised learning task) besides the main task during  training, and update the pre-trained model using test data with the (self-supervised) test-time objective before the final prediction. 
\citet{sun2020test} uses rotation prediction as the self-supervised task. Later, TTT++ \citep{liu2021ttt++} considers contrastive loss as the self-supervised task in addition to aligning the features by comparing the statistics of the source data with those of the current test batch. Recently, \citet{2023tttflow} minimizes the distribution shift, between the train and test distributions, estimated using normalizing flows. \citet{ttt_mae} shows that using masked autoencoding (MAE) \citep{he2022masked} as the self-supervised task for TTT achieves substantial improvements in image recognition under various distributional shifts. TTT-based techniques are applied to other domains such as videos~\citep{azimi2022self, wang2023testtime}, natural language processing (NLP)~\citep{banerjee2021self} and compressed sensing of medical images \citep{ttt_mri}, where the self-supervised task varies across domains.
In this work, we extend TTT framework to speech domain for the first time.
Our work extends the TTT-MAE framework \citep{ttt_mae} to speech by using  work on audio MAE trained with spectrograms \citep{mae_audio_neurips, mae_audio_2}.

\textbf{Parameter efficient fine-tuning (PEFT).}
Fine-tuning entire pre-trained models achieves state-of-the-art performance for various downstream tasks \citep{kenton2019bert, raffel2020exploring, ssl_speech_review}. However, as the size of these models increases rapidly, updating the models in parameter-efficient ways becomes crucial~\citep{param_eff_class, 2022adaptformer}. In the NLP domain, PEFT techniques typically refer to either: a) insertion of new learnable modules with fewer parameters as compared to the whole model; or b) modifying carefully selected parameters of the model. In \cite{param_efficient_finetune, pfeiffer2020mad, sung2022vl, li2021prefix}, only the additional parameters added to the pre-trained models are fine-tuned for the downstream tasks while ~\cite{param_eff_class, 2022bitfit} fine-tune a subset of parameters for downstream tasks without inserting any new modules. \citet{2022bitfit} show that just fine-tuning the bias parameters, which constitute about 0.1\% of the overall parameters, can outperform full fine-tuning, especially for small datasets.
Motivated from NLP, PEFT techniques are also applied to pre-trained models trained using images \citep{param_Eff_vision, 2022adaptformer, adapters_vision, lee2022surgical}, and speech~\citep{speechTranslation_param, Param_eff_asr_proj}. In this work, we study PEFT techniques applied to speech in the context of TTT, focusing on the following questions: 1) Can PEFT techniques achieve comparable or improved performance compared to full fine-tuning in TTT? 2) Can PEFT be more stable than full fine-tuning in TTT?

\textbf{Self-Supervision using masking in speech.} Denoising autoencoders~\citep{vincent2008extracting} is one of the earliest forms of self-supervision in speech \citep{lu2012speech}. Subsequent advancements in self-supervision have predominantly focused on masked language modeling (MLM) \citep{liu2020mockingjay, baevski2020wav2vec, hsu2021hubert, gong2022ssast, iclr_HuangZY0022}. Many of these works employ transformer networks trained with MLM frameworks to achieve state-of-the-art performance on various speech-related downstream tasks. More recently, the concept of a masked autoencoder (MAE) based on the Vision Transformer (ViT) has been extended to the audio and speech domain \citep{mae_audio_neurips, mae_audio, mae_audio_2, mae_audio_3}. In MAE, only the non-masked spectrogram patches are encoded, distinguishing it from other approaches that encode both masked and non-masked input (wave/spectrogram) segments for self-supervised pre-training. This distinction makes MAE-based models computationally appealing. 
In this work, we analyze the effect of distributional shift on the performance of MAE for speech and use TTT-based approach to enhance the robustness of MAE  in the presence of such distributional shifts.

\section{Method}
\label{method}

\textbf{Pre-training MAE.} Our masked autoencoder (MAE) for speech, following \citet{mae_audio_neurips, mae_audio_2}, aims to reconstruct the masked patches of the speech Mel-spectrogram with an asymmetrical encoder-decoder architecture. Below, we provide a brief overview of the MAE.

First, we transform the input speech waveform into 128-dimensional Mel-spectrograms using a Hanning window of size 25ms for every 10ms. Next, we divide the spectrogram into a sequence of non-overlapping patches, each patch sized 16 $\times$ 16.
These patches are then flattened and embedded with a linear projection layer. To provide positional information, fixed sinusoidal positional embeddings are added to the embedded patches. Afterward, we randomly mask 75\% of the patches while preserving  positional indices of all the patches. This enables the decoder to reconstruct the spectrogram. For the encoder, only  unmasked patches are used to generate latent representations. The decoder then tries to reconstruct the original spectrogram, given  latent representations of the encoder and  masked patches as input. The latent representations and masked patches are organized in the initial order before being provided as input to the decoder.
During training, the objective is to minimize mean squared error (MSE) between reconstructed and input spectrograms, averaged over the masked patches.

\textbf{Train-time training.} For the downstream tasks, we only use the encoder and discard the decoder. The latent representations generated by the encoder are provided as input to the task-specific classifier head. In this paper, we consider three different ways to use the pre-trained encoder for downstream tasks: 1) Linear probing: freeze the encoder to use it as a feature extractor and train only the classifier head; 2) Fine-tuning: train both encoder and classifier head, end-to-end, for the downstream task; 3) Linear-probing and fine-tuning (LP-FT): first train only the classifier head using linear probing, and then fine-tune both encoder and classifier head end-to end as explained in \citep{LPFT2022}.


\textbf{Test-time training.}
Similar to \citep{sun2020test, ttt_mae}, we use a Y-shaped architecture: a shared encoder network $e$ followed by two heads, a self-supervised head $g$ and a task-specific classifier head $h$. Here, $e$ and $g$ are the encoder and decoder networks of the pre-trained MAE, respectively.
The classifier head $h$, uses a linear projection from the dimension of the encoder features to the number of classes, depending on the downstream task.

\begin{figure}
  \centering
  \includegraphics[width=0.86\linewidth]{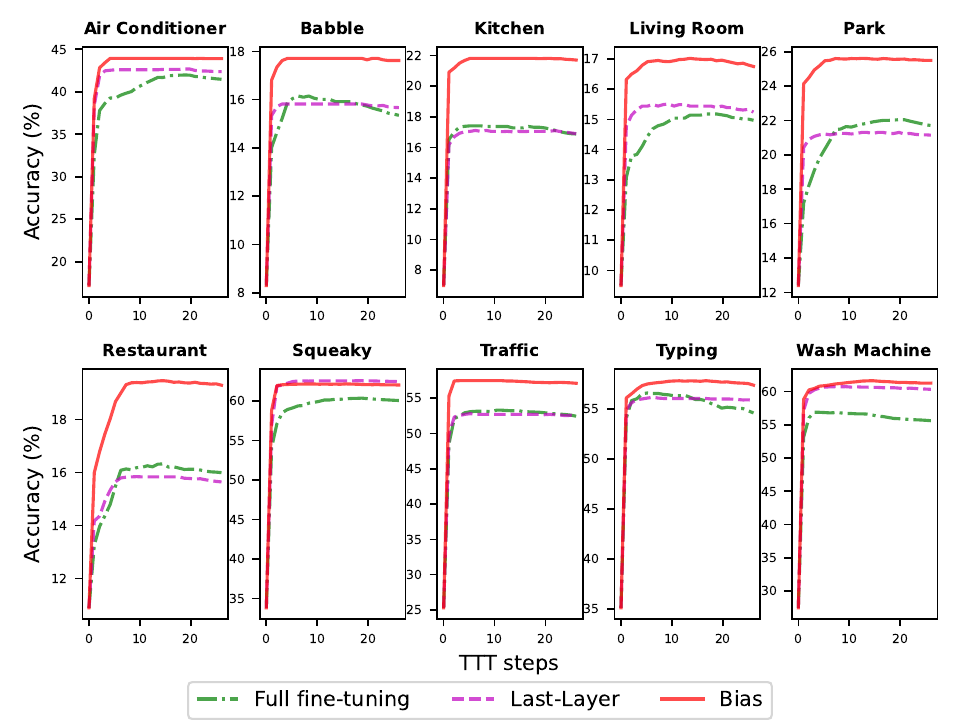}
  \caption{\small We compare the accuracy (speaker identification) across TTT steps between three different variants of TTT (full, last-layer and bias fine-tuning). For most distributional shifts, full fine-tuning shows degradation in performance with longer test-time training (after 20 steps) whereas bias and last layer fine-tuning show relatively stable performance even after 25 steps}
  \label{fig:ttt_steps}
\end{figure}

When using TTT, we use linear-probing to update the weights of the classifier head $h$ and freeze the weights of the shared encoder $e$ during train-time training. As explained in \cite{ttt_mae}, we found that linear-probing is more suitable for TTT as compared to full-model fine-tuning. 
At test-time, the parameters of the shared encoder are updated to minimize the self-supervised loss. We explore the following approaches to update the weights of the encoder during test-time:

1) \textbf{Full fine-tuning}: In full fine-tuning, all parameters of the shared encoder are updated to minimize the self-supervised loss across various augmentations of a single test sample. However, the large size of pre-trained models, such as the MAE used in this study with 75M parameters, makes full fine-tuning computationally expensive during test time. Moreover, extensive steps of full fine-tuning during test time, as shown in Figure \ref{fig:ttt_steps}, can result in performance degradation. Since there is no validation data available at test time, early stopping is not an option.

Therefore, it is desirable to maintain stability in performance during Test-Time Training (TTT). However, even when using the SGD optimizer as suggested in \citep{ttt_mae}, full fine-tuning still exhibits performance degradation in speech-related tasks. Furthermore, TTT techniques entail a higher computational cost as TTT is approached as a one-sample learning problem. This means that the model can only be adapted to one test sample at a time, and batch processing is not feasible under the assumption that each test sample is subject to a different distribution shift.

\begin{wraptable}[12]{R}{0.5\linewidth}
\vspace{-0.3cm}
\small
\caption{\small Trainable parameters in different blocks of our MAE for speech. Number of Parameters in one Encoder block are same as the number of parameters in each of the first, middle or last blocks of the encoder.}
\label{tab:params}
\centering
\begin{tabular}{lr}
\toprule
 & \#parameters(Proportion \%) \\
 \midrule
MAE & 74751488 (100.00) \\
Encoder & 64457216 (86.23) \\
Decoder & 10294272 (13.77) \\
One Encoder block & 7087872 (9.48) \\
Bias - MAE & 96000 (0.13) \\
Bias - Encoder & 77568 (0.10) \\
\bottomrule
\end{tabular}
\end{wraptable}

To address these challenges, we investigate parameter-efficient fine-tuning techniques (PEFT) that have proven to be highly effective in supervised learning tasks within the field of NLP. However, their application in the context of TTT has not previously been explored.

2) \textbf{Parameter-efficient fine-tuning (PEFT)}: Here we explore different PEFT techniques to adapt the encoder to the test sample during inference. We conduct experiments using four different PEFT techniques for TTT: (i) First block: updating only the final/last block of the encoder, (ii) Last block: updating only the final/last block of the encoder, (iii) Middle block: updating only the middle block of the encoder, and (iv) Bias: updating only bias parameters of the encoder using the self-supervised task on the test sample.
As shown in Table \ref{tab:params}, the number of trainable parameters for updating any one block (first, middle, or last) accounts for 9.48\% of the total parameters, while updating the bias parameters involves just 0.1\% of the total parameters. In this study, we primarily focus on bias fine-tuning for TTT due to the following reasons:
\begin{itemize}[leftmargin=0.2cm,nolistsep,topsep=0pt]
\item Bias fine-tuning is much more lightweight compared to full fine-tuning, training 830 times fewer parameters (0.78M vs. 64M).
\item \citet{2022bitfit} demonstrated that, in supervised learning with limited training data, fine-tuning only bias parameters yields superior performance to full fine-tuning. Since TTT-MAE can be regarded as a one-sample unsupervised domain adaptation technique, we investigate effectiveness of updating only bias parameters during test-time training.
\item TTT techniques incur higher computational costs as TTT is approached as a one-sample learning problem. This means that the model can only be adapted to one test sample at a time, and batch processing is not feasible when each test sample is drawn from a different distribution. We show that bias fine-tuning allows processing an entire batch of test samples.
\end{itemize}

\textbf{Distributional shifts in speech.}
In this work, we identify and examine two types of distributional shifts: (1) those resulting from background noise that degrades/distorts speech quality, and (2) natural distributional shifts resulting from inter-speaker variations, including gender, age, and speaking style.

\begin{figure}
  \centering
  \includegraphics[width=1\linewidth]{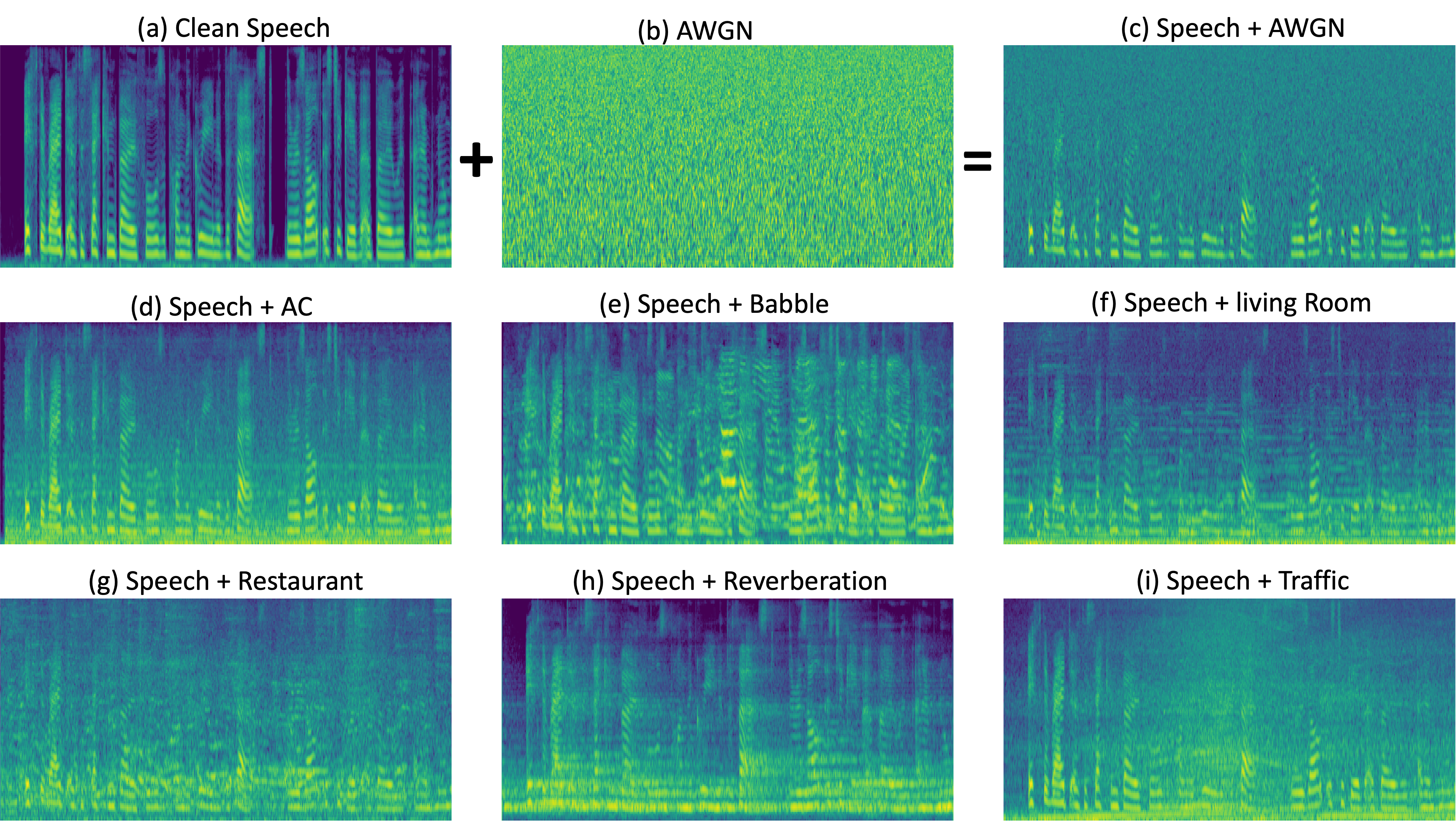}
  \caption{\small Mel-spectrograms of speech with different distributional shifts due to background noises added at 0 dB SNR. (c) shows the distribution shift in Mel-spectrogram of clean speech (see (a)) when added with AWGN ((b)). (d)-(i) shows the Mel-spectrograms of speech when added with different background noises. Characteristics of background noises (e)-(i) vary with time thus distorting the characteristics of speech critical for speech-based applications. For instance, panel (e) shows how babble noise, which has similar characteristics to speech, when added to speech,  introduces patterns similar to clean speech along both time and frequency dimensions, and thus distorts the patterns in clean speech. Similarly, living room (see (f) and restaurant (see (g)) noises distort speech patterns in time and frequency dimensions.}
  \label{fig:specgram}
\end{figure}

To generate degraded speech, we introduce background noise to clean speech signals at a specified signal-to-noise ratio (SNR) (see Figure \ref{fig:specgram}). We identify two categories of background noise: (1) \textit{Time-invariant}, where noise characteristics remain constant over time. Examples include additive white Gaussian noise (AWGN) and air conditioner (AC). (2) \textit{Time-varying}, where noise characteristics change over time. Examples include background babble, living room, restaurant, reverberation, and traffic. Distributional shifts introduced by these types of noise are generally difficult to learn, even with adversarial training. 
Furthermore, some noises (e.g. babble, restaurant, and reverberation) exhibit patterns similar to speech, which can corrupt and contaminate  information contained in the original signal, such as linguistic content, speaker characteristics, emotions, etc. 

In this study, we explore  significance 
of TTT in handling distributional shifts introduced by both background noise and natural variations in speech. We show that TTT-based methods consistently outperform non-TTT techniques with significant margins. Use of PEFT techniques, particularly BitFit, further improve the performance and stability of TTT under different distributional shifts.
To show the effectiveness of TTT under different distributional shifts, we conduct the following experiments: 1) Training with clean speech and testing with speech corrupted by various background noises. 2) Training models on one speaking style and testing with another speaking style. 3) Training with speech from one gender and testing with the other gender. 4) Training with either younger or older speakers and testing with the other.

\section{Experiments and Results}
\label{Sec: Exp_res}

\textbf{Implementation Details}
In all experiments, we use 9-layer ViT by default as the MAE encoder. For the decoder, we use a 3-layer Transformer. We use Voxceleb2 dataset \citep{chung2018voxceleb2} for pre-training the MAE. We pre-train the MAE for 120K steps with a batch size of 392, which takes about 7 days using 8 Nvidia RTX6000 24GB GPUs. The AdamW optimizer with an initial learning rate of 0.001 and a weight decay of 0.05 is applied. The learning rate has a cosine decay schedule \citep{chen2020cosine} with 20K warmup steps. 
We transform raw waveform (mono-channel sampled at 16 KHz) into 128 Mel-frequency bands extracted with 20 ms Hanning window and 10 ms stride. We use the original speech spectrograms and apply no augmentations. We only use random masking of the spectrograms as an augmentation, with a masking ratio of 0.75 for pre-training.

During TTT, for each test sample, we train only the encoder (freezing the decoder weights)for $20$ steps using SGD optimizer with a fixed learning rate of 2.5e-3, batch size of 128, momentum of 0.9 and weight decay of 0.2. We also show performance plots for $25$ steps of TTT (see Figure \ref{fig:ttt_steps}). During TTT, we follow the same procedure as pre-training: mask 75\% of the input patches and provide the unmasked patches as input to the encoder whereas all the patches are provided to the decoder. Then update the encoder weights using reconstruction loss (MSE) on the masked patches as the objective function. We follow the same procedure for full, bias, first-layer, middle-layer and last layer fine-tuning.s
Moreover, we do not use any augmentation on top of random masking for TTT. We performed most of these experiments using a single Nvidia A40 48GB GPU. Unless  and otherwise specified, we report results for TTT after 20 TTT steps.

\begin{table}[ht]
\small 
  \caption{\small Details of the datasets used for pre-training and downstream tasks. Total Duration is in hours and Average Length is in seconds. The tasks Speaker ID, Emotion Recog., Ltd Vocab. ASR refer to Speaker identification, emotion recognition and limited vocabulary ASR, respectively.}
  \label{tab:data}
  \centering
  \begin{tabular}{llrrrrr}
  \toprule
  Dataset & Task & \#Classes & \#Speakers & \#Total & Total &Average \\
 & & & & Samples & Duration & Length \\
  \midrule
  VoxCeleb2& Pre-training & -- & 5994 & 770000 & 2300 & 8.6 \\
  VCTK & Speaker ID & 109 & 109 & 42075 & 44 & 3.8 \\
  CREMA-D & Emotion Recog. & 4 & 91 & 4397 & 3.04 & 2.5 \\
  IEMOCAP & Emotion Recog. & 4 & 10 & 5531 & 7.00 & 4.6 \\
  RAVDESS & Emotion Recog. & 4 & 24 & 672 & 0.70 & 3.7 \\
  TESS & Emotion Recog. & 4 & 2 & 1600 & 0.92 & 2.1 \\
  Speech Commands & Ltd. Vocab. ASR & 12 & 2618  & 105829 & 29.5 & 1.0 \\
  \bottomrule
  \end{tabular}
\end{table}


\textbf{Dataset details.} Table \ref{tab:data} provide details of the datasets used in this work. We perform speaker identification using VCTK~\citep{vctk}; emotion recognition using CREMA-D~\citep{cao2014crema}, IEMOCAP~\citep{busso2008iemocap}, RAVDESS~\citep{livingstone2018ryerson} and TESS~\citep{dupuis2011recognition} datasets; low-vocabulary speech recognition using Speech commands~\citep{warden2018speech} dataset.
We use original speech samples from each of the datasets in the pre-training and  train-time training phases. To evaluate models' performance under different distributional shifts, we introduce diverse background noises sourced from Microsoft's Scalable Noisy Speech Dataset (MS-SNSD) \citep{reddy2019scalable}. These noises are exclusively added during the testing phase and are not used in pre-training or train-time training of the models.

\begin{figure}[ht]
 \centering
 \begin{subfigure}[b]{0.48\textwidth}
 \centering
 \includegraphics[width=\textwidth]{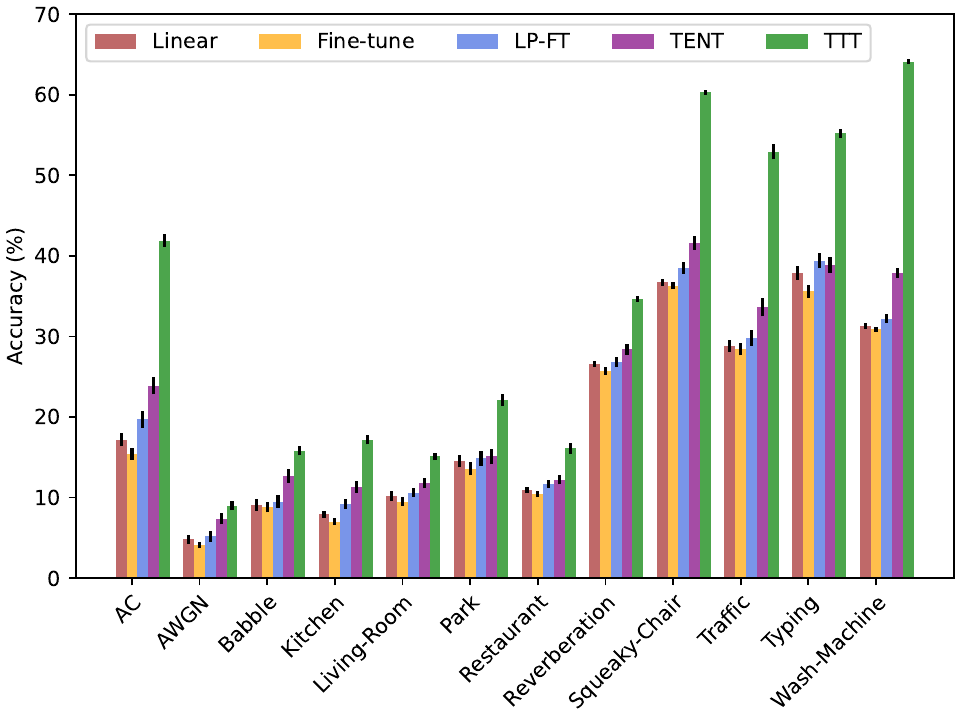}
 \caption{Speaker identification (Accuracy (\%)) on VCTK}
 \label{fig:p_vctk}
 \end{subfigure}
 \hfill
 \begin{subfigure}[b]{0.48\textwidth}
 \centering
 \includegraphics[width=\textwidth]{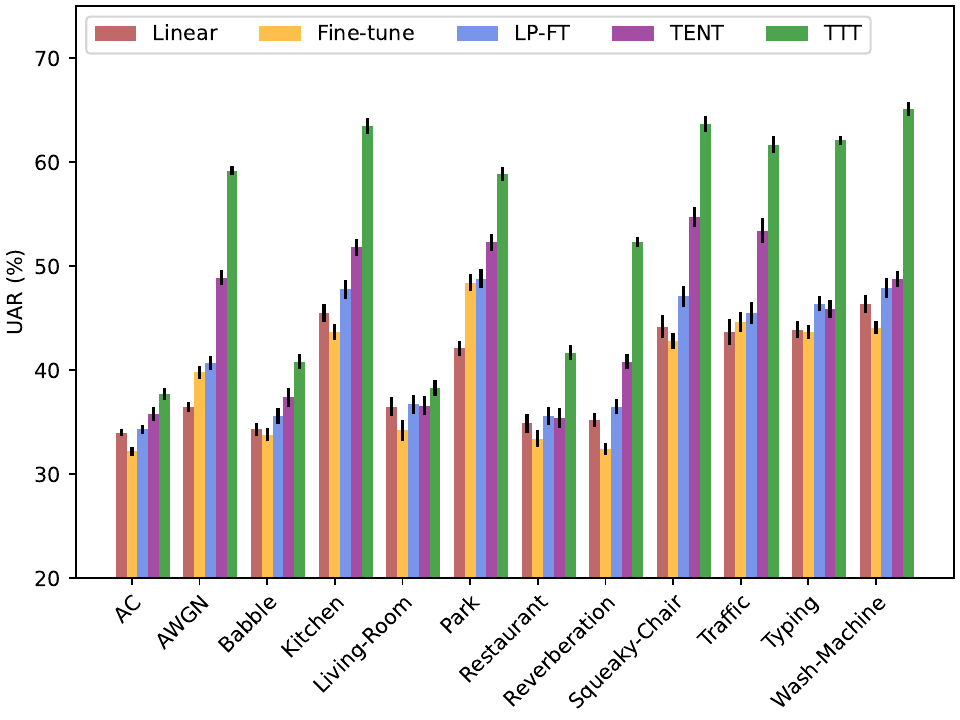}
 \caption{Emotion classification (UAR(\%)) on Crema-D}
 \label{fig:p_crem}
 \end{subfigure}
  \caption{\small Compare TTT with non-TTT approaches (linear probing, fine-tuning, LP-FT, TENT) under different different shifts due to background noises (noises added at 0 dB). TTT significantly outperforms the non-TTT approaches across all the distributional shifts. Results averaged over 3 runs. Emotion classification reported in terms of unweighted average recall (UAR (\%))}
  \label{fig:perf_ttt}
\end{figure}

\textbf{Test-time Training vs No Test-time Training.} In Figure \ref{fig:perf_ttt}, we compare TTT for speech with different non-TTT techniques (linear probing, fine-tuning, LP-FT~\citep{LPFT2022} and TENT~\citep{wang2021tent}) for different background noises unseen during training. TTT outperforms the non-TTT techniques for every unseen background noise condition. When comparing between non-TTT techniques, for most of the background noises, simple linear-probing performs better than fine-tuning. Similar to images in \citep{LPFT2022}, LP-FT performs better than both linear-probing and fine-tuning.
TENT, a test-time adaptation technique, performs better or comparable to LP-FT but performs inferior to TTT. This can be attributed to the fact that TENT requires a large set of test samples to learn the distribution of the test sample but performs poorly under one-sample testing condition, as illustrated in \citep{khurana2021sita}.

Even though TTT achieves significant improvements in performance under different distributional shifts, there are a few shortcomings of TTT when applied to speech such as high memory requirements during TTT, degradation in performance across TTT steps and inability to process batch of test samples, as shown in Figure \ref{fig:ttt_steps} and explained in Section~\ref{method}.

We overcome these issues by incorporating PEFT techniques into TTT.
PEFT techniques are light weight as we need to fine-tune fewer parameters compared to full fine-tuning, thus requiring lesser memory (see Table~\ref{tab:params}).
We also find that, for speech, PEFT techniques achieve better consistency in performance compared to full fine-tuning across the TTT steps (see Figure~\ref{fig:ttt_steps}).
Here we compare  performance of different PEFT techniques used during TTT under different distributional shifts in Table \ref{tab:diff_ttt}.
PEFT techniques of fine-tuning a specific block performs comparable to full-fine-tuning. Similar to \citep{lee2022surgical}, we find  certain layers are more suitable for a specific set of background noises and there is no single block which is optimal for every background noise. Selecting a layer to fine-tune during TTT is not feasible as we only have a single test sample. To overcome this issue, we use Bitfit \citep{2022bitfit} for TTT, where we fine-tune only the bias parameters of the encoder during TTT. Bitfit, a light-weighted fine-tuning approach, consistently performs better (comparable) to full-fine-tuning across all the distributional shifts.


\begin{table}
\small
\caption{\small Performance of different variants of TTT under distributional shifts due to background noises added at 0dB SNR. Different variants of TTT: Full refers to Full fine-tuning; First, Middle, Last and Bias refer to fine-tuning only the first layer, middle layer, last layer and bias parameters, respectively. AWGN $\rightarrow$ additive white Gaussian noise. Bias fine-tuning performs better than other variants of TTT across most of the distribution shifts due to background noise}
  \label{tab:diff_ttt}
\begin{tabular}{lrrrrrrrrrr}
\toprule
 & \multicolumn{5}{c}{Speaker Identification} & \multicolumn{5}{c}{Emotion Recognition} \\ 
\cmidrule(lr){2-6}  \cmidrule(lr){7-11} \vspace{-0.3cm} \\ 
 & Full & First & Middle & Last & Bias & Full & First & Middle & Last & Bias \\
\midrule
Air Conditioner & 41.8 & 42.6 & 40.8 & 42.5 & \textbf{43.9} & 37.7 & 37.6 & 36.1 & 38.0 & \textbf{42.8}  \\
Gaussian (AWGN) & 8.8 & 9.1 & 8.2 & 9.3 & \textbf{10.4} & 59.2 & 59.0 & 58.6 & 59.7 & \textbf{62.8} \\
Babble & 15.7 & 17.0 & 16.0 & 15.7 & \textbf{17.3} & 40.8 & 42.3 & 41.6 & 41.2 & \textbf{42.8}  \\
Kitchen & 17.2 & 17.8 & 16.9 & 17.1 & \textbf{21.8} & 63.5 & 64.1 & 63.7 & 64.6 & \textbf{65.9}  \\
Living Room & 15.1 & 16.3 & 16.0 & 15.4 & \textbf{16.9} & 38.3 & 39.4 & 38.5 & \textbf{40.5} & 40.3  \\
Park & 22.1 & 22.4 & 23.3 & 21.2 & \textbf{25.6} & 58.9 & 59.6 & 58.2 & 60.1 & \textbf{62.1}  \\
Restaurant & 16.1 & 18.5 & 16.3 & 17.2 & \textbf{19.4} & 41.7 & 43.1 & 41.8 & 42.5 & \textbf{43.9}  \\
Reverberation & 34.7 & 35.4 & 35.2 & 35.8 & \textbf{37.9} & 52.3 & 53.4 & 52.2 & 52.5 & \textbf{54.7}  \\
Squeaky-Chair & 60.3 & 60.6 & 60.8 & \textbf{62.5} & 62.1 & 63.7 & 64.3 & 64.4 & 63.6 & \textbf{65.5}  \\
Traffic & 52.9 & 55.9 & 54.3 & 52.7 & \textbf{57.2} & 61.7 & 65.3 & 63.2 & 65.7 & \textbf{66.8}  \\
Typing & 55.2 & 57.4 & 57.4 & 55.9 & \textbf{57.7} & 62.1 & 63.2 & 59.7 & 63.3 & \textbf{64.6}  \\
Washing-Machine & 56.0 & 60.8 & 60.5 & 60.4 & \textbf{61.4} & 64.1 & 65.1 & 65.0 & 65.4 & \textbf{66.2}  \\
\bottomrule
\end{tabular}
\end{table}

\begin{table}[!htbp]
\caption{\small Emotion recognition under natural distributional shifts caused by (a) \textbf{speaking style variations}: Train model using CREMA-D (CRE) dataset and test with other datasets i.,e IEMOCAP (IEM), RAVDESS (RAV) and TESS. Column CRE: matched condition -- train and test on CREMA-D dataset, (b) \textbf{Gender variations}: Train model using speech data from speakers of one gender and test with speakers from other gender. F-M: train on female and test on male; M-F: train on male and test on female; F-F and M-M: train and test speakers from the same gender. We use CREMA-D dataset for these experiments. TTT variants (Full and Bias fine-tuning) outperform non-TTT methods (linear probing and fine-tuning) across different natural distribution shifts.}
\label{tab:nat_ds}
\begin{subtable}{0.50\textwidth}
\centering
\small
\caption{evaluate across datasets - variation in speaking style}
\label{tab:diff_data}
\begin{tabular}{lrrrr}
\toprule
& \multicolumn{1}{l}{CRE} & \multicolumn{1}{l}{IEM} & \multicolumn{1}{l}{RAV} & \multicolumn{1}{l}{TESS} \\
\midrule
Linear & 52.2 & 35.4 & 34.7 & 33.1 \\
Fine-tune & 61.5 & 32.5 & 33.8 & 34.3 \\
\midrule
Full & 66.8 & 45.5 & \textbf{44.3} & 39.4 \\
Bias & \textbf{67.3} & \textbf{45.9} & 44.2 & \textbf{40.6} \\
\bottomrule
\end{tabular}
\end{subtable}
\begin{subtable}{0.47\textwidth}
\small
\centering
\caption{Across genders - variation in gender}
\label{tab:diff_gen}
\begin{tabular}{lrrrr}
\toprule
 & F-F & F-M & M-M & M-F \\
\midrule
Linear & 54.2 & 31.6 & 52.1 & 30.9 \\
Fine-tune & 59.4 & 29.8 & 58.6 & 30.2 \\
\midrule
Full & 66.0 & 54.1 & 61.5 & 56.8 \\
Bias & \textbf{68.2} & \textbf{56.5} & \textbf{64.0} & \textbf{60.5} \\
\bottomrule
\end{tabular}
\end{subtable}
\end{table}

\begin{wraptable}[14]{R}{0.5\linewidth}
\centering
\small
\caption{\small Emotion classification under age variation. We use TESS dataset consisting of two female speakers: one Younger (Y) and one older (O). We train on one speaker and test using other speaker's speech. Y-O $\rightarrow$ train on Y and test with O; O-Y $\rightarrow$ train on O and test with Y}
\label{tab:diff_age}
\begin{tabular}{lrrrr}
\toprule
& Y-Y & Y-O & O-O & O-Y \\
\midrule
Linear & 49.3  & 29.8  & 50.2  & 31.7  \\
Fine-tune & 51.8  & 28.2  & 52.3  & 30.4  \\
\midrule
Full  & 52.2  & 40.8  & 52.5  & 41.6  \\
Bias & \textbf{52.4} & \textbf{41.9} & \textbf{52.9} & \textbf{42.8} \\
\bottomrule
\end{tabular}
\end{wraptable}

\textbf{Evaluation under natural distributional shifts.} We evaluate  performance of TTT techniques across different natural distributional shifts caused by inter-speaker variations, e.g., speaking style (Table \ref{tab:diff_data}), gender (Table \ref{tab:diff_gen}), and age (Table \ref{tab:diff_age}).

In Table \ref{tab:diff_data}, we compare performance of non-TTT (Linear probing and fine-tuning) with TTT (full fine-tuning (Full) and Bias) techniques for distributional shift due to speaking style variation. Here we train on CREMA-D  (emotively acted utterances in American English) and test with samples from IEMOCAP (emotive conversations enacted in American English), RAVDESS (emotively acted utterances in North American English) and TESS (emotively acted utterances in Canadian English). TTT-based techniques achieve significantly better performance compared to non-TTT methods for speaking style variations.

To evaluate the performance of TTT in addressing distributional shifts caused by gender variations for the task of emotion classification (CREMA-D dataset), we conducted training using speech data from one gender (either female or male) and carried out testing using speech data from the opposite gender (male or female). Table \ref{tab:diff_gen} provides a comparison between non-TTT and TTT-based techniques. 
For gender variations, non-TTT approaches exhibit a steep decline in performance between matched (F-F and M-M) and mismatched (F-M and M-F) conditions. In contrast, TTT approaches (full and Bias) show very small degradation in performance between matched and mismatched conditions. 
For both matched and mismatched conditions, TTT with Bias fine-tuning performs better than full fine-tuning.

We evaluate  performance of TTT under age variations using TESS dataset for the task of emotion recognition (refer to Table~\ref{tab:diff_age}). The TESS dataset was collected from two female speakers: a younger (Y) speaker aged 26 years and an older speaker aged 64 years. In the O-Y scenario, we trained the model using younger (Y) speaker's speech and tested it with older (O) speaker's speech. Similarly, in the Y-O scenario, we trained with the older speaker (O) and tested with the younger (Y) speaker. For unmatched train-test conditions (Y-O and O-Y), TTT techniques outperformed non-TTT techniques, with bias fine-tuning performing better than full fine-tuning. Interestingly, even in matched conditions (Y-Y and O-O), TTT-based techniques performed comparably or better than non-TTT techniques for all the natural distributional shifts.



\begin{wrapfigure}[34]{r}{0.5\textwidth}
    \centering
    \vspace{-0.2in}
    \includegraphics[width=\linewidth]{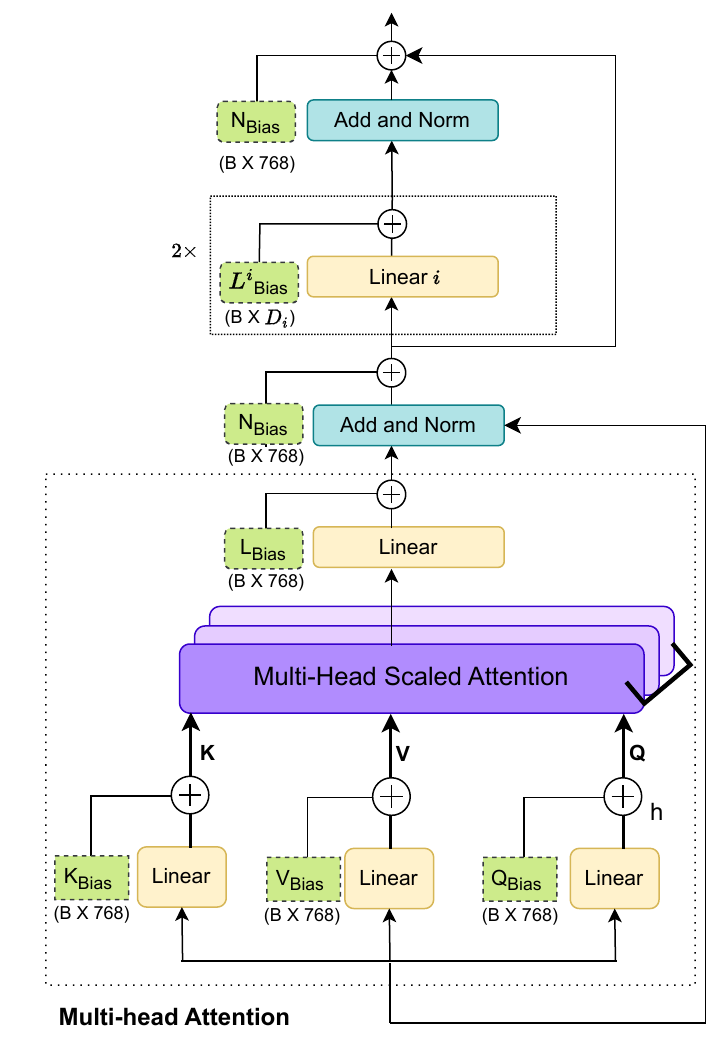}
    \caption{\small Illustration of TTT-Bias Fine-tuning on Batch of Test Samples: for each example in a batch containing $B$ samples, we create a learnable parameter corresponding to each trainable bias as shown in green boxes and the outputs of the modules are adjusted accordingly as shown. Only the parameters in green boxes need to be fine-tuned and since these parameters are not shared across examples in a batch, TTT-bias can take advantage of GPU batch processing to improve throughput. }
    \label{fig:bt_ttt}
\end{wrapfigure}

\textbf{Improved utilization of computational resources.} We discuss an approach to process a batch of test samples using TTT for the real world scenario of each test sample associated with a different distributional shift. Since each test sample gets its own copy of parameters, TTT can be applied with only one test sample at a time. However, bias fine-tuning gives us an unique opportunity to apply TTT to a batch of examples while still ensuring that each sample gets its own copy of parameters. For example, consider a simple linear layer containing weight matrix $W$ and a bias-vector $b$; given a batch ${\bf x}$ containing $B$ samples, the output of this layer can be written as ${\bf y} = W{\bf x} + b$. In TTT-bias, each example in this batch gets its own $b$ while sharing $W$; now, suppose that the linear layer gets an additional input $\Delta b$ (initialized to zeros) that contains $B$ learnable vectors and is used to compute the output ${\bf y} = W{\bf x} + b + \Delta b$: since $\Delta b$ contains $B$ learnable parameters, each parameter can be optimized independently while taking advantage of the GPU batch-processing. We illustrate this in Figure \ref{fig:bt_ttt}: in practice, this can be implemented as forward hooks in PyTorch without changing the model code.
Refer to the supplementary material for further analysis and results.

\section{Conclusion}
In this work, we extend test-time training to speech related tasks such as speaker identification and emotion recognition. 
We extend the TTT-MAE, proposed in \cite{ttt_mae} for image recognition, to improve the performance of speech-related downstream tasks under a variety of distributional shifts. In our application of TTT-MAE to speech, we observed that TTT is sensitive to hyperparameters such as training steps and subset of parameters considered for optimization. To overcome these issues,  we applied PEFT techniques to make TTT more stable and scalabale. We find that PEFT techniques, being light-weight, achieve better or comparable performance to full fine-tuning. Specifically, we find that bias fine-tuning, motivated from BitFit, improves both performance and stability. Further, we propose a new approach to process batches of test samples using bias fine-tuning for TTT.




\bibliographystyle{abbrvnat}
\bibliography{references}


\section*{Limitations}
In this work, we evaluate test-time training techniques for speech applications. We consider parameter-efficient fine-tuning alternatives and identify that bias fine-tuning makes the test-time training method robust to optimization hyperparameters while also improving inference speed. Nonetheless, test-time training of large pre-trained models requires appropriate compute resources such as GPUs for faster inference. While we demonstrate that bias fine-tuning allows for batch processing of test-examples and increases throughput, inference with TTT is still not as scalable as non-TTT inference. 

\section*{Broader Impact}
In this work, we introduce TTT for speech and show that TTT improves performance of various speech-related downstream tasks under different distributional shifts relative to the training data. This includes distributional shifts caused due to speaking style, gender and age based differences -- see Table \ref{tab:nat_ds} for speaking style and gender variations, and Table \ref{tab:diff_age} for age variations. Acquiring labels such that the dataset is unbiased can be prohibitively expensive or simply unfeasible. For example, biases can get introduced into training datasets because the examples do not accurately represent all possible variations in speaker demographics such as gender, age, accent, geographical, cultural and economical background; our results show that TTT can be used to adapt models trained on biased datasets. 
Like any classification technology that infers personal information (e.g. speaker detection), such technologies come with broad potential risks and benefits depending on the downstream applications. For instance, with improved performance of machine learning models, privacy of personal information might either be at increased risk, or better protected depending on how the technology is used. As another, particularly relevant, example, our own work in this area is driven to support our medical collaborations with health care practitioners, where there is a great need and value to mitigate the effects of distributional shift between training data and the test data, to de-bias the system for more effective diagnostic-support use.

\newpage
\section*{Dataset Details}
\textbf{Voxceleb2 Dataset:} In this work, We pre-trained the MAE models using the Voxceleb2 dataset~\citep{chung2018voxceleb2}. Voxceleb2 dataset contains 1.1M video clips collected from over 6000 celebrities, extracted from 150K unique videos uploaded to Youtube. Average duration of the video clips is around 8 seconds. For this work, we only considered video clips with at least 3 seconds in duration, resulting  in a total of 1M video clips to pre-train the models.

\textbf{Voice Cloning ToolKit (VCTK) Dataset:} The VCTK dataset~\citep{vctk} consists of approximately 42,000 clean audio clips uttered by 109 native English speakers with various accents. Each speaker reads out about 400 sentences. The average length of each audio clip is around 4 seconds and the total duration of the audio clips is approximately 44 hours. In this paper, we use the VCTK dataset for the task of speaker identification. We randomly split the utterances spoken by each speaker into 3 splits: 80\% training set, 10\% validation set and remaining 10\% as test set. We have a total of 33662 training samples, 4206 validation samples and 4206 test samples.

\textbf{Crowd-Sourced Emotional Multimodal Actors Dataset (CREMA-D):} CREMA-D~\citep{cao2014crema} is a crowd sourced audiovisual dataset consisting of emotive utterances collected from 91 speakers (48 male actors and 43 female actresses). Actors spoke from a group of twelve sentences. In this work, we used data corresponding to four emotional categories: anger, happy, sad and neutral,  resulting in 4900 utterances. Total durations of the utterances is 3.4 hours with an average length of 2.5 seconds.

\textbf{Interactive Emotional Dyadic Motion Capture (IEMOCAP)}: IEMOCAP~\cite{busso2008iemocap} consists of scripted and improvised dialogues by 10 speakers. It is composed of 5 recording sessions, each including speech from one male speaker and one female speaker. In this work, we used 4 emotional classes: anger, happy, sad and neutral. Following previous works \citep{fayek2017evaluating, pepino2021emotion}, we relabeled excitement samples as happy. This resulted in a total of 5531 utterances with a total duration of 7 hours.

\textbf{Ryerson Audio-Visual Database of Emotional Speech and Song (RAVDESS):} RAVDESS dataset consists of audio-visual recordings from twelve male and twelve female actors with North American accent reciting English sentences. In this work, we used utterances with 4 emotional classes (anger, happy, sad and neutral) resulting in 672 utterances with a total duration of 0.7 hours.

\textbf{Toronto Emotional Speech Set (TESS):} TESS \citep{dupuis2011recognition} dataset consists of 200 target words spoken in the carrier phrase “Say the word \_\_\_” by two actresses, aged 26 and 64, in different emotive states. In this work, we used utterances spoken in 4 different emotions i.e., anger, happy, sad and neutral. We have a total of 1600 utterances, 800 by each actress, with a total duration of 0.92 hours and an average length of 2.1 seconds.

\textbf{Speech Commands V2 (SPCV2) Dataset:} The Speech Command V2 \citep{warden2018speech} dataset consists of 105,829 1-second recordings of 35 common speech commands. In \citep{warden2018speech}, the 35 commands are segregated into 11 different classes – 10 common commands are selected as 10 different classes and the remaining 25 commands are considered as one class. The silence audio samples were created from background noise sounds, which are not used as a class in SPCV2. The training/validation/testing set has 84,843/9,981/11,005 1-second recordings, respectively.

\textbf{Additional Rationale for Dataset Choices} 
Our reasons for choosing the datasets VCTK, CREMA-D and speech commands for analysis are: 1) Due to the vast diversity of speakers, both male and female speakers. 2) Clean speech: Recorded in a soundproof environment. For the evaluation of TTT, we wanted the original recordings to contain as little background noise as possible. This is particularly important for the evaluation of the systems in presence of different background noises which should not be present in the training data. 

\section*{Batch Processing in TTT}
\begin{figure}[ht]
    \centering
    \includegraphics[width=0.92\linewidth]{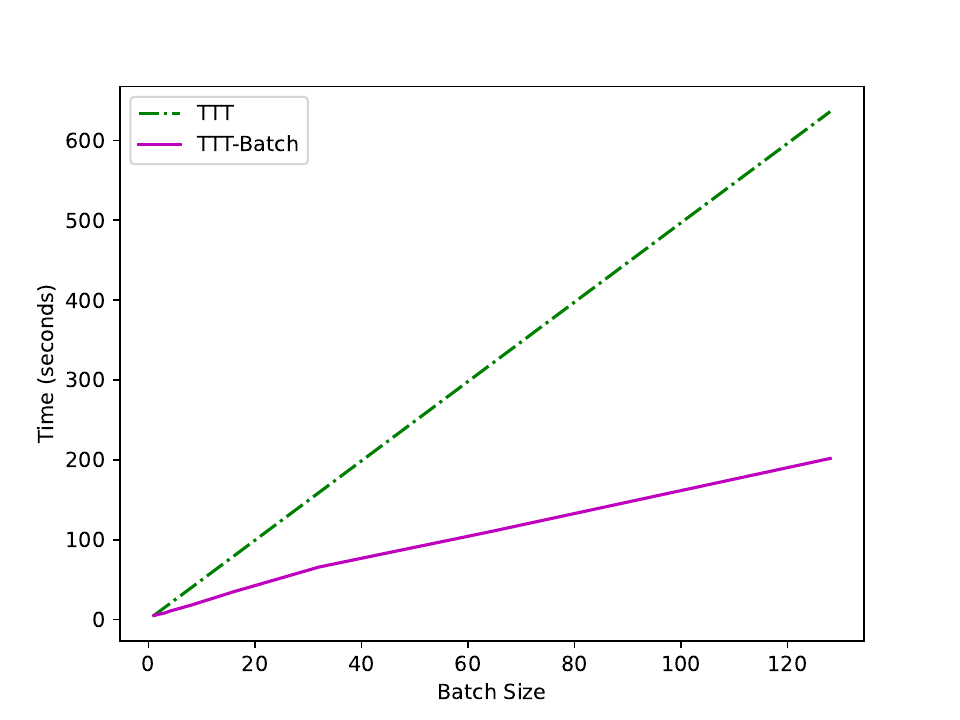}
    \caption{\small Timing analysis of TTT-Bias Fine-tuning on Batch of Test Samples. Here we provide the timing analysis of the approach shown in Figure \ref{fig:bt_ttt} to process a batch of test samples. We vary the batch size of the test samples and compare the time taken by  TTT (process only one sample at a time) and TTT-Batch (process batch of test samples simultaneously. We show that the difference in time taken by TTT-Batch and TTT increases as the batch size increases). For the purpose of this illustration, we use 32 augmentations for each TTT step}
    \label{fig:time_anal_bat}
\end{figure}

Figure~\ref{fig:time_anal_bat} shows the timing analysis for processing a batch of test samples using (1) TTT (process single sample at a time), (2) TTT-Batch: Process a batch of test samples using the approach illustrated in Figure~\ref{fig:bt_ttt}. As shown in Figure~\ref{fig:time_anal_bat}, as the batch size increases, the throughput of TTT-Batch improves.

\section*{Results on Speech Commands Dataset}
In Table~\ref{tab:spc}, we compare the performance of non-TTT techniques (Linear, Fine-tune) with TTT-based techniques (TTT-Full, TTT-Bias) using speech commands dataset for the task of low-vocabulary speech recognition. Results show that TTT techniques outperform non-TTT techniques under different distributional shifts. Also in TTT-based techniques, the Bias fine-tuning performs better (comparable) to full fine-tuning across all the distributional shifts.

\begin{table}[!htbp]
\caption{Results on Speech Commands dataset for the task of low-vocabulary speech recognition. performance in terms of Accuracy (in \% averaged over 3 random seeds) when tested with samples corrupted with different noises. All noises added at 0dB SNR. Results show that the TTT-based techniques significantly improves the performance over non-TTT techniques under different distributional shifts. Bias fine-tuning performs better than full-fine-tuning across all distributional shifts.}
\label{tab:spc}
\centering
\begin{tabular}{lcccccccccc}
\toprule
  & {\rotatebox[origin=c]{50}{Clean}} & {\rotatebox[origin=c]{50}{AC}} & {\rotatebox[origin=c]{50}{Babble}} & {\rotatebox[origin=c]{50}{Kitchen}} & {\rotatebox[origin=c]{50}{Living}} & {\rotatebox[origin=c]{50}{Park}}  & {\rotatebox[origin=c]{50}{Restaurant}} & {\rotatebox[origin=c]{50}{Reverb}} & {\rotatebox[origin=c]{50}{Traffic}} & {\rotatebox[origin=c]{50}{Typing}} \\
\midrule
Linear & 92.3 & 51.4 & 30.9 & 54.2 & 35.8 & 50.2 & 37.8 & 40.7 & 56.1 & 53.8 \\
 Fine-tune  & 95.2 & 50.9 & 29.3 & 53.6 & 34.2 & 49.3 & 38.3 & 39.2 & 55.4 & 54.5 \\
\midrule
TTT-Full & \textbf{95.4} & 62.2 & 37.4 & 63.3 & 45.5 & 58.3 & 45.5 & 55.1 & 63.7 & 64.1 \\
TTT-Bias & 94.8 & \textbf{63.7} & \textbf{39.2} & \textbf{64.4} & \textbf{46.4} & \textbf{60.4}  & \textbf{47.8} & \textbf{57.2} & \textbf{66.4} & \textbf{67.3} \\
\bottomrule
\end{tabular}
\end{table}

\section*{Comparison with Supervised Methods}
In this section, we will compare the performance of MAE for speech and MAE + TTT with models such as CNN, LSTM and CNN-LSTM, trained using supervised techniques from scratch. CNN model consists of 3 convolutional layers with filters (256, 128, 128) followed by a feed-forward layer. LSTM network consists of 2 LSTM layers each with 128 ReLU units followed by a feed-forward layer. CNN-LSTM model consists of 3 convolutional layers followed by a single LSTM layer consisting of 64 units. All the supervised models (CNN, LSTM and CHNN-LSTM) are trained using Mel-spectrograms. Same approach as explained in Section~\ref{Sec: Exp_res} is followed to extract the Mel-spectrograms. 
Tables \ref{tab:crema_sup} and \ref{tab:vctk_sup} provide the compare supervised techniques (CNN, LSTM, CNN-LSTM) with pre-trained models (MAE (Fine-tune), TTT-Full and TTT-Bias) for CREMA-D and VCTK datasets, respectively. Pre-trained models perform better than the supervised models trained from scratch on clean speech and speech corrupted with different background noises. Similar to MAE (Fine-tune), supervised techniques show huge degradation in performance when tested with speech corrupted with different background noises. TTT-based techniques improve performance over supervised and pre-trained models when tested with noisy speech.

\begin{table}[!htbp]
\caption{Comparison with supervised techniques (CNN, LSTM and CNN-LSTM). Emotion recognition results in terms of UAR (in \% averaged over 3 random seeds) using the models trained on Crema dataset and tested with samples corrupted with different noises. All noises added at 0dB SNR. Pre-trained models perform better than supervised models trained from scratch on both, clean and speech corrupted with background noises. When tested with speech degraded with background noise, TTT-based techniques outperform  both supervised and pre-trained models}
\label{tab:crema_sup}
\centering
\begin{tabular}{lcccccccc}
\toprule
  & Clean & Babble & Kitchen & \begin{tabular}[c]{@{}r@{}}Living\\ Room\end{tabular} & Park  & Restaurant & Traffic & Typing \\
\midrule
CNN & 59.8 & 29.5 & 35.6 & 30.8 & 44.6 & 32.5 & 40.7 & 41.0 \\
LSTM & 58.7 & 30.2 & 34.9 & 31.1 & 44.9 & 31.8 & 39.9 & 41.2 \\
CNN-LSTM  & 60.8 & 31.6 & 35.8 & 30.9 & 45.3 & 33.2 & 40.6 & 41.4 \\
\midrule
 MAE (Fine-tune) & 61.5  & 33.6 & 43.7 & 34.2 & 48.4 & 33.5 & 44.6 & 43.7 \\
\midrule
TTT-Full  & 66.8  & 40.8 & 64.1 & 38.3 & 58.9 & 41.7 & 61.7 &  62.1 \\
TTT-Bias & \textbf{67.3} & \textbf{42.8} & \textbf{65.9} & \textbf{40.3} & \textbf{62.1}  & \textbf{43.9} & \textbf{66.8} & \textbf{64.6} \\
\bottomrule
\end{tabular}
\end{table}

\begin{table}[!htbp]
\caption{Comparison with supervised techniques (CNN, LSTM and CNN-LSTM). Speaker identification results in terms of Accuracy (in \% averaged over 3 random seeds) using the models trained on VCTK dataset and tested with samples corrupted with different noises. All noises added at 0dB SNR. Pre-trained models perform better than supervised models trained from scratch on both, clean and speech corrupted with background noises. When tested with speech degraded with background noise, TTT-based techniques outperform  both supervised and pre-trained models}
\label{tab:vctk_sup}
\centering
\begin{tabular}{lcccccccc}
\toprule
  & Clean & Babble & Kitchen & \begin{tabular}[c]{@{}r@{}}Living\\ Room\end{tabular} & Park  & Restaurant & Traffic & Typing \\
\midrule
CNN & 90.5 & 7.9 & 6.6 & 9.2 & 11.5 & 8.6 & 24.6 & 31.8 \\
LSTM & 91.1 & 8.2 & 6.5 & 8.8 & 11.8 & 8.7 & 24.8 & 32.2 \\
CNN-LSTM & 91.6 & 8.4 & 6.8 & 9.1 & 11.7 & 8.7 & 25.3 & 32.5  \\
\midrule
 MAE (Fine-tune) & \textbf{92.1} & 8.8 & 7.0 & 9.5 & 13.6 & 10.4 & 28.4 & 35.6 \\
\midrule
TTT-Full  & 90.3 & 15.7 & 17.2 & 15.1 & 22.1 & 16.1 & 52.9 & 55.2  \\
TTT-Bias & 91.4 & \textbf{17.3} & \textbf{21.8} & \textbf{16.9} & \textbf{25.6}  & \textbf{19.4} & \textbf{57.2} & \textbf{57.7} \\
\bottomrule
\end{tabular}
\end{table}

\newpage
\section*{Speaking Style Variation: IEMOCAP Dataset}
Table~\ref{tab:iemo_spk_sty} shows the effect of speaking style variation on the performance of models trained with IEMOCAP dataset (See Table~\ref{tab:diff_data} for effect of speaking style variation on CREMA-D dataset). Here we train on IEMOCAP (emotive conversations enacted in American English)
and test with samples from CREMA-D (emotively acted utterances in American English),
RAVDESS (emotively acted utterances in North American English) and TESS (emotively acted
utterances in Canadian English). TTT-based techniques achieve significantly better performance compared to non-TTT methods for speaking style variations. 

\begin{table}[!htbp]
\caption{Speaking style variation based distribution shift on IEMOCAP dataset: Emotion recognition results in terms of UAR (in \% averaged over 3 random seeds) using models trained on IEMOCAP dataset and tested with samples from other datasets. IEMOCAP consists of conversational speech recording whereas CREMA-D consists of read speech with enacted emotions, RAVDESS is emotive read speech collected from speakers with north American accent and Tess dataset consists of read speech recordings collected from Canadian speakers. TTT-based techniques (TTT-Full and TTT-Bias) outperform non-TTT techniques under the distributional shift caused due to speaking style variation}
\label{tab:iemo_spk_sty}
\centering
\begin{tabular}{lrrrr}
\toprule
 & IEMOCAP & CREMA & RAVDESS & TESS \\
 \midrule
Linear & 56.3 & 37.4 & 31.3 & 35.6 \\
Finetune & 60.8  & 40.5  & 30.6  & 32.9 \\
\midrule
TTT-Full  & \textbf{63.3}  & 49.3  & 42.1  & 44.6 \\
TTT-Bias & 63.2 & \textbf{51.4} & \textbf{42.8} & \textbf{45.5} \\
\bottomrule
\end{tabular}
\end{table}

\newpage
\section*{Distributional Shifts in Speech}
\begin{figure}[!htbp]
 \centering
 \begin{subfigure}[b]{1.0\textwidth}
 \centering
 \includegraphics[width=\textwidth]{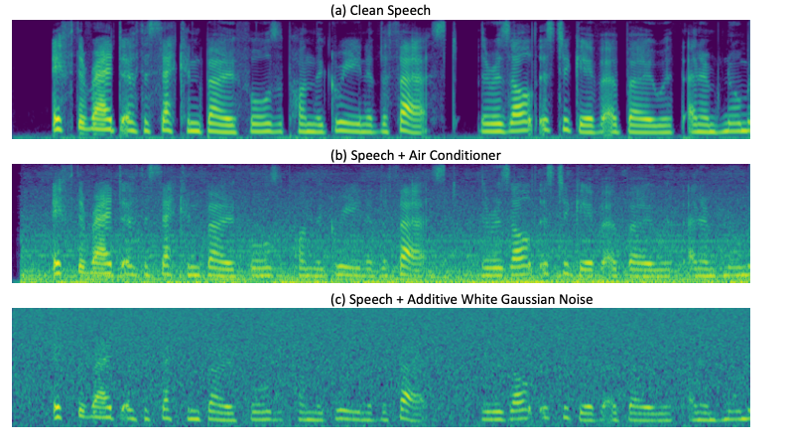}
 \label{fig:spec_1}
 \end{subfigure}
 \hfill
 \begin{subfigure}[b]{1.0\textwidth}
 \centering
 \vspace{-0.5cm}
 \includegraphics[width=\textwidth]{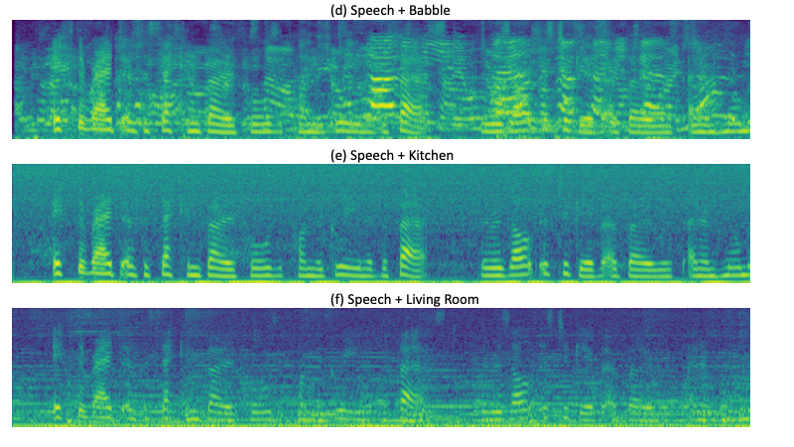}
 \label{fig:spec_2}
 \end{subfigure}
  \caption{\small Mel-spectrograms of speech with distributional shifts caused due to different background noises added at 0 dB SNR}
  \label{fig:spec_dist}
\end{figure}

\begin{figure}[!htbp]
 \centering
 \begin{subfigure}[b]{1.0\textwidth}
 \centering
 \includegraphics[width=\textwidth]{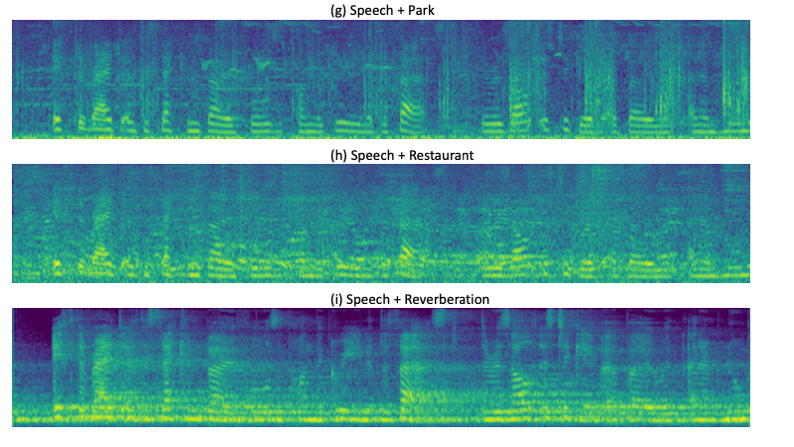}
 \label{fig:spec_3}
 \end{subfigure}
 \hfill
 \begin{subfigure}[b]{1.0\textwidth}
 \centering
 \vspace{-0.5cm}
 \includegraphics[width=\textwidth]{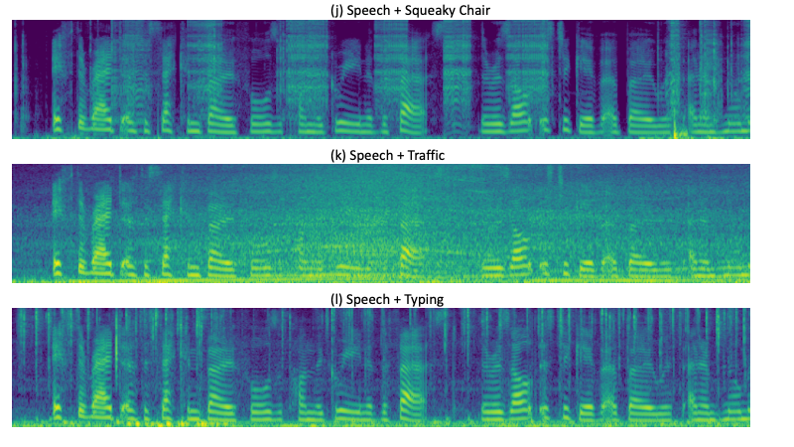}
 \label{fig:spec_4}
 \end{subfigure}
 \hfill
 \begin{subfigure}[b]{1.0\textwidth}
 \centering
 \vspace{-0.5cm}
 \includegraphics[width=\textwidth]{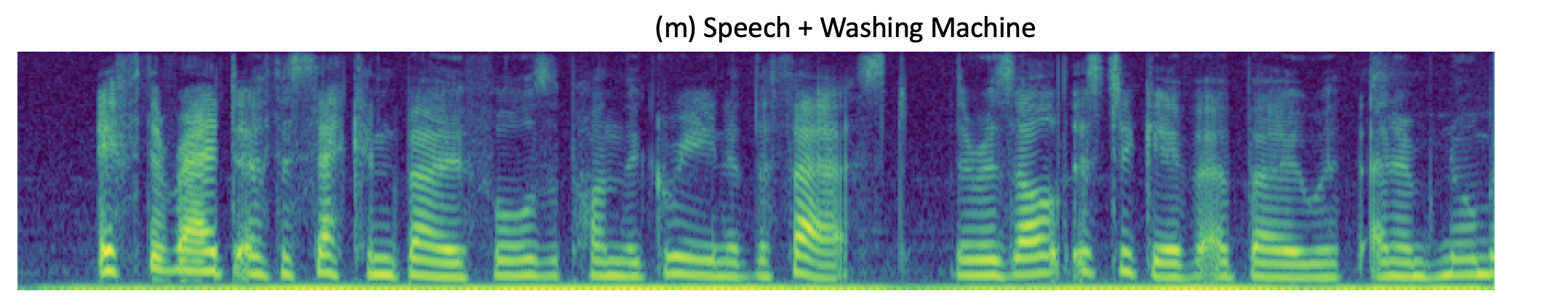}
 \label{fig:spec_5}
 \end{subfigure}
  \caption{\small Mel-spectrograms of speech with distributional shifts caused due to different background noises added at 0 dB SNR}
  \label{fig:spec_dist1}
\end{figure}

In Figures \ref{fig:spec_dist} and \ref{fig:spec_dist1}, we show the spectrograms of the clean speech distorted with different background noises, added at 0 dB SNR. 

\newpage
\thispagestyle{empty}

\begin{table}
\small
\caption*{\small Table 1: Performance comparison of different variants of TTT. Additional adapter units are trained in two ways (1) Adap-1: train adapter units during both train-time training and during TTT; (2) Adap-2: train adapter units only during TTT. Adap-2 performs better than Adap-1 for most of the distribution shifts. Bias fine-tuning performs better than adapter units (Adap-1 and Adap-2) for most of the distribution shifts}
  \label{tab:diff_ttt_adap}
  \centering
\begin{tabular}{lrrrrrrrr}
\toprule
 & \multicolumn{4}{c}{Speaker Identification} & \multicolumn{4}{c}{Emotion Recognition} \\ 
\cmidrule(lr){2-5}  \cmidrule(lr){6-9} \vspace{-0.3cm} \\ 
 & Full & Adap-1 & Adap-2 & Bias & Full & Adap-1 & Adap-2 & Bias \\
\midrule
Clean & 90.3 & 90.2 & 90.7 & 91.4 & 66.8 & 66.0 & 67.1 & 67.3 \\
\midrule
Babble & 15.7 & 14.7  & 16.5 & \textbf{17.3} & 40.8 & 38.5 & 40.3 & \textbf{42.8}  \\
Kitchen & 17.2 & 18.2 & 20.5 & \textbf{21.8} & 63.5 & 63.1 & 64.2 & \textbf{65.9}  \\
Living Room & 15.1 & 14.5 & 16.0 & \textbf{16.9} & 38.3 & 37.1 & 38.5 & \textbf{40.3}  \\
Park & 22.1 & 17.8 & 21.9 & \textbf{25.6} & 58.9 & 56.3 & 57.7 & \textbf{62.1}  \\
Restaurant & 16.1 & 14.7 & 17.6 & \textbf{19.4} & 41.7 & 39.2 & 42.7 & \textbf{43.9}  \\
Reverberation & 34.7 & 33.1 & 35.4 & \textbf{37.9} & 52.3 & 49.1 & 51.8  & \textbf{54.7}  \\
Squeaky-Chair & 60.3 & 60.1 & \textbf{62.3} & 62.1 & 63.7 & 63.4 & 65.3 & \textbf{65.5}  \\
Traffic & 52.9 & 47.2 & 50.6 & \textbf{57.2} & 61.7 & 59.2 & 60.8 & \textbf{66.8}  \\
Typing & 55.2 & 55.1 & \textbf{57.8} & 57.7 & 62.1 & 61.9 & 64.2 & \textbf{64.6}  \\
\bottomrule
\end{tabular}
\end{table}



\begin{table}
\small
\caption*{\small Table 2: Performance comparison of TTT-Bias (Bias) with other SSL techniques. TTT with bias fine-tuning performs better than other SSL techniques without TTT. W2V-2.0, M. Jay and USAT refer to the models Wav2vec-2.0, Mockingjay and UniSpeech-SAT, respectively.}
  \label{tab:diff_ssl}
  \centering
\begin{tabular}{lrrrrrrrrrr}
\toprule
 & \multicolumn{5}{c}{Speaker Identification} & \multicolumn{5}{c}{Emotion Recognition} \\ 
\cmidrule(lr){2-6}  \cmidrule(lr){7-11} \vspace{-0.3cm} \\ 
 & W2V-2.0 & M.Jay & Hubert & USAT & Bias & W2V-2.0 & M.Jay & Hubert & USAT & Bias \\
\midrule
clean & 92.0 & 87.8 & 92.3 & \textbf{93.8} & 91.4 & 61.3 & 57.4 & 61.9 & 63.1 & \textbf{67.3}  \\
\midrule
Babble & 9.2 & 8.6 & 9.5 & 10.9 & \textbf{17.3} & 34.1 & 32.7 & 34.9 & 36.1 & \textbf{42.8}  \\
Kitchen & 7.3 & 6.8 & 7.4 & 9.2 & \textbf{21.8} & 45.1 & 43.4 & 44.8 & 47.5 & \textbf{65.9}  \\
Living Room & 9.7 & 9.4 & 10.1 & 12.4 & \textbf{16.9} & 34.7 & 33.5 & 35.1 & 36.4 & \textbf{40.3}  \\
Park & 14.2 & 13.1 & 14.8 & 16.3 & \textbf{25.6} & 49.6 & 47.9 & 50.2 & 52.7 & \textbf{62.1}  \\
Restaurant & 10.6 & 9.2 & 10.9 & 12.0 & \textbf{19.4} & 34.5 & 32.2 & 35.2 & 37.4 & \textbf{43.9}  \\
Reverberation & 26.4 & 24.8 & 26.2 & 27.3 & \textbf{37.9} & 37.6 & 35.4 & 38.1 & 40.1 & \textbf{54.7}  \\
Squeaky-Chair & 36.2 & 35.8 & 36.9 & 38.5 & \textbf{62.1} & 42.8 & 40.9 & 43.0 & 44.6 & \textbf{65.5}  \\
Traffic & 29.1 & 28.0 & 28.8 & 30.4 & \textbf{57.2} & 44.7 & 43.5 & 45.1 & 45.4 & \textbf{66.8}  \\
Typing & 35.5 & 34.9 & 35.8 & 37.2 & \textbf{57.7} & 44.3 & 43.2 & 44.8 & 45.2 & \textbf{64.6}  \\
\bottomrule
\end{tabular}
\end{table}

\begin{table}[!htbp]
\caption*{\small Table 3: Effect of masking ratio during TTT. Speaker identification results in terms of Accuracy using the models trained on VCTK dataset and tested with samples corrupted with different noises. During TTT, masking ratio of 75\% performs better than (comparable to) other ratios. We chose 75\% as it reduces the computational cost and achieves top performance}
\label{tab:mask_rat}
\centering
\begin{tabular}{lcccccccc}
\toprule
Ratio & Babble & Kitchen & \begin{tabular}[c]{@{}r@{}}Living\\ Room\end{tabular} & Park  & Restaurant & Traffic & Typing & Avg. \\
\midrule
\midrule
25\% & 15.3 & 18.9 & 14.9 & 21.8 & 18.1 & 51.4 & 52.5 & 27.55  \\
50\% & 17.1 & 21.2 & 17.0 & 25.2 & 19.6 & 56.5 & 57.4 & 30.54  \\
75\% & 17.3 & 21.8 & 16.9 & 25.6 & 19.4 & 57.2 & 57.7 & 30.85\\
90\% & 14.2 & 16.2 & 13.8 & 18.7 & 15.1 & 48.4 & 49.3 & 25.1  \\
\bottomrule
\end{tabular}
\end{table}

\end{document}